\documentclass[journal=jpccck,manuscript=article]{achemso}

\usepackage[english]{babel}
\usepackage{amsmath}
\usepackage{graphicx}
\usepackage{esint}
\usepackage{epsfig}
\usepackage{color}

\def\beq{\begin{equation}}
\def\eeq{\end{equation}}
\def\beqa{\begin{eqnarray}}
\def\eeqa{\end{eqnarray}}

\def\half{{\ss 1\over 2}}

\def\cH{{\mathcal H}}
\def\cL{{\mathcal L}}

\def\ss{\scriptstyle}

\def\ss{\scriptstyle}

\author{Yonatan Dubi$^1,^2$}
\email{jdubi@bgu.ac.il}
\affiliation{$^1$Department of Chemistry and $^2$Ilse-Katz Institute for Nanoscale Science and Technology, Ben-Gurion University of the Negev, Beer-Sheva 84105, Israel}

\title{Interplay between Dephasing and Geometry and Directed Heat Flow in Exciton Transfer Complexes}
\date{\today}
\keywords{Exciton transfer network, directed heat transport}
\begin{document}  
\begin{abstract}
 The striking efficiency of energy transfer in natural photosynthetic systems and the recent evidence of long-lived quantum coherence in biological light harvesting complexes has triggered 
 much excitement, due to the evocative possibility that these systems - essential to practically all life on earth -- use quantum mechanical effects to achieve optimal functionality. 
 A large body of theoretical work has addressed the role of local environments in determining the transport properties of excitons in photosynthetic networks and the survival of quantum coherence in a 
 classical environment. Nonetheless, understanding the connection between quantum coherence, exciton network geometry and energy transfer efficiency remains a challenge. Here we address this connection from the perspective of heat transfer within the exciton network. Using a non-equilibrium open quantum system approach and focusing on the Fenna-Matthews-Olson complex, we demonstrate that finite local dephasing can 
 be beneficial to the overall power output. The mechanism for this enhancement of power output is identified as a gentle balance between quantum and classical contributions to the local heat flow, such that the total heat flow is directed along the shortest paths and dissipation is minimized. Strongly related to the spatial network structure of the exciton transfer complex, this mechanism elucidates how energy flows in photosyntetic excitonic complexes. \end{abstract}


\section{Introduction} 
Green plants, algae and some bacteria exploit photosynthesis to convert solar energy into chemical energy, making sugars out of water and carbon dioxide and releasing oxygen. Since 
practically all life on Earth depends on plants for food and on oxygen for breathing, photosynthesis can therefore be viewed as a fundamental processes that has shaped life on earth as we know 
it \cite{Nelson2005,Blankenship2013}. Despite differences in the photosynthetic machinery of different species, a good model for many photosynthetic complexes - whether in simple unicellular algae or in complex higher plants - is a general three-part structure comprising the antenna, the transfer network, and the reaction center. The antenna captures energy from sunlight and subsequently excites the electrons of the pigment (e.g., chlorophyll in green plants) from their ground state. The excited electrons, which combine with holes to make excitons, travel from the antenna through an 
intermediate protein exciton transfer complex  (ETC) to the reaction center where, in a simplified view of a complex process \cite{Blankenship2013,Jordanides2001}, they participate in the chemical reaction that generates oxygen. 

Across species, transfer complexes, key determinants of photosynthetic process efficiency, vary both in size - e.g., from as small as seven chromophores in green sulfur bacteria to tens and 
even hundreds of nodes in higher plants - and in their network structure and complexity \cite{Blankenship2013,Oijen1999}. The simplest complex, for example, is probably the Fenna-Matthews-Olson (FMO) complex \cite{Fenna1974,Camara-Artigas2003,Adolphs2006,Cho2005} in green sulfur 
bacteria. The FMO complex consists of a triad of molecular sub-units, each of which is a network of seven interconnected chromophores (schematically depicted in Fig.~\ref{fig1}a, arranged in two weakly-connected branches (sub-networks) that are separately connected to the antenna (sites 1 and 6 in Fig.~\ref{fig1}a) and jointly connected to the reaction center (via site 3 in the Fig.~\ref{fig1}a).

Remarkably, ultrafast nonlinear spectroscopy experiments with transfer complexes have provided evidence that excitons within the ETC exhibit quantum coherence in the form of long-lasting oscillations in the ETC's two-dimensional spectra \cite{Lee2007,Engel2007a,Collini2010,Calhoun2009,Panitchayangkoon2011,Panitchayangkoon2010}. The exciting possibility that photosynthetic systems optimize their performance via quantum effects is redefining our understanding of photosynthesis and has established an unexpected connection between physics, chemistry and biology. In addition, photosynthetic processes - and specifically the exciton transfer - are extremely efficient \cite{Blankenship2011,Blankenship2013}. Elucidating the mechanisms underlying that efficiency may translate into design principles that will promote the development of better artificial bio-insipired energy harvesting and energy-transmitting technologies \cite{Odobel2013,Scholes2011a,Eisenberg2014}.

Not surprisingly, the evocative results from these experiments have stimulated a large body of theoretical research into the nature of excitonic transfer in ETCs, and the role of the environment -- as a source of decoherence -- in determining the exciton transfer efficiency (broadly reviewed in  Refs.~\cite{Ishizaki2012,Fleming2011,Cheng2009,Dawlaty2012,Collini2013,Lambert2013,Scholes2005,Pachon2012,Scholes2014} and many references therein). While many theories have proposed mechanisms by which the environment can assist exciton transfer across the ETC \cite{Rebentrost2009,Mohseni2008,Caruso2009,Chin2012,Chin2013,Plenio2008,Kassal2012,Caruso2014,Li2015,Marais2013,Sinayskiy2012,Chen2015,Ghosh2011,Ghosh2011a,Chen2013}, the question of 
whether natural systems indeed maintain quantum coherence and utilize it to enhance their photosynthetic performance is still open \cite{Kassal2013,Cheng2009,Tiersch2012,Ishizaki2012,Miller2012,Ritschel2011,Romero2014,Dawlaty2012,Li2012}. Furthermore, the vast majority of the theoretical studies address the 
ETC efficiency by defining it as the number of excitons that, upon excitation from a coherent source, can cross the transfer network. Although this is a natural way to evaluate the transfer efficiency, the relation of this approach to the fundamental role of the ETCs as {\sl energy transmitters} is not straightforward. Theoretical studies which did take into account power output (e.g., \cite{Xu2014,Dorfman2013}) 
did not take the transfer network geometry into account. The nature of the interplay between power output, optimal performance, dephasing and network geometry remains and open question \cite{Levi2015}. 

In this article, therefore, we adopt a different approach. Using a non-equilibrium open quantum systems approach we directly calculate the 
heat flux that flows into the ETC from the antenna and outward to the reaction center, which allows for natural definitions for thermodynamic efficiency and performance of the photosynthetic exciton transfer process. We discuss the FMO complex (indeed, the "fruit-fly" of photosynthetic ETC studies), but out results hold for more general complexes (as discussed also in the supplementary information). We find that although thermodynamic efficiency consistently decreases with any increase in environment-induced local dephasing (characterized by a dephasing time), both the power absorbed by the ETC from the antenna and that transferred to the reaction center increase in the presence of local dephasing (compared to the dephasing-free situation), and exhibit a maximum at optimal dephasing times. 

By calculating of the local heat currents inside the ETC network, the mechanism for this optimal heat transfer is 
identified as a balance between the quantum- and classical behavior, leading to the directed flow of heat across the ETC. This mechanism is clarified by an example that can be solved analytically. In addition, we analyze the effect of a localized vibrational mode on the heat transport, and find that in contrast to the local dephasing, interaction with the local mode always impedes 
the power output. Finally, we discuss the durability of the FMO transfer network structure (vis-a-vis its power output) with respect to significant structural variations. Taken together, our results elucidate the relation between quantum coherence, dephasing and the network structure of the photosynthetic ETC. 

\section{Model and method}
\subsection{Hamiltonian and quantum master equation}
Here we introduce the key elements of our calculations. The FMO exciton transfer network is parameterized by the FMO Hamiltonian  \cite{Cho2005}, 
\beq
\cH_{FMO}=\sum^7_{i=1}\epsilon_i c^{\dagger}_i c_i+\sum_{i,j} t_{i,j}c^{\dagger}_i c_j~~,
\eeq
where $c^{\dagger}_i,~c_i$ are the creation and annihilation operators of an exciton at chromophore $i$, respectively, $\epsilon_i$ are the excitation energies and $t_{i,j}$ the inter-chromophore couplings (the eighth chromophore is ignored here \cite{SchmidtamBusch2010,Tronrud2009,Yeh2014}, but is considered in the supplementary information).
$\cH_{FMO}$ corresponds to the graph in Fig. 1a: on-site excitation energies are marked next to the molecules, and interchromophore couplings are marked by arrows, such that the extent of arrow thickness and darkness correlates with stronger couplings. Within the model, energy is continuously injected into the ETC from the antenna (red arrows, sites 1 and 6 in Fig.~1a) at a rate of $\gamma_{\mathrm{inj}}$, and extracted to the reaction center (red arrow, site 3 
in Fig.~1a) at a rate of $\gamma_{\mathrm{out}}\approx 10 \mathrm{ ps}^{-1}$ \cite{Guan2013,Plenio2008} (taken as a constant hereafter). Both injection (excitation) and extraction processes are incoherent and continuous \cite{Manzano2013,Pelzer2014} (a pulse excitation is considered in the supplementary information). 

In addition to its interaction with the source and the sink, the exciton can lose its phase coherence at each site at a dephasing rate of  $\gamma_{\mathrm{deph}}$. Described using the Lindblad quantum master equation \cite{Breuer2002,DiVentra2008,Lindblad1976}, the loss of phase coherence can be interpreted alternatively (from the stochastic Schroedinger equation   \cite{Breuer2002,DiVentra2008})  as a measurement randomly performed on the exciton by the local vibrations with an average time of  $\tau_{\mathrm{deph}}=\gamma^{-1}_{\mathrm{deph}}$.

The Lindblad quantum master equation is given by 
\beq
\dot{\rho}=-i [\cH,\rho]+\cL [\rho],
\eeq
where $\rho$ is the density matrix of the excitonic system, 
$ \cH $
is the FMO Hamiltonian  and the Lindbladian is
$
\cL[\rho]=-\frac{1}{2}\{V^{\dagger}V,\rho \}+V \rho V^{\dagger}~~. 
$
Here $[\cdot,\cdot]~~,\{ \cdot,\cdot \} $ are the commutation and anti-commutation relations, respectively, and $V$ constitute the Lindblad operators, that encode the operation of the environment on the system. The source and sink terms are defined through the $V$-operators 
$ V_{source}=\gamma_{\mathrm{inj}}^{-1/2}c^\dagger_1,\gamma_{\mathrm{inj}}^{-1/2}c^\dagger_6$ and $V_{sink}=\gamma_{\mathrm{out}}^{-1/2} c_3 $.
The local dephasing is defined with the operators $V_{{\mathrm{deph}},i}=\gamma_{\mathrm{deph}}^{-1/2} c^{\dagger}_i c_i$. Similarly, the exciton recombination is taken into account here with a Lindblad terms $V^{+}_{rec,i}=(\gamma_{rec} f(\epsilon_i))^{1/2}c^\dagger_i$ and  $V^{-}_{rec,i}=(\gamma_{rec}(1-f(\epsilon_i)))^{1/2}c_i$, where $f(\epsilon)$ is the Fermi-Dirac distribution function. Such a form ensures detailed balance at equilibrium \cite{Ajisaka2015}. However, because the recombination rate is much smaller than the dephasing processes, it is not important in the calculations shown here.   

For the case of a localized vibration, an additional term is added to the Hamiltonian, 
\beq \cH_{vib}=\hbar \omega_0 \sum_i a^\dagger_i a_i +\lambda \sum_i(a_i+a^\dagger_i) c^\dagger_i c_i ~~,\eeq
 such that $\lambda$ is an 
exciton-vibration interaction term. The interaction of the localized vibration with the ambient environment is addressed via a bosonic Linblad $V$-operators of the form $V^{+}_{phn,i}=(\gamma_{\mathrm{phn}} n_B(\omega_0))^{1/2} a^{\dagger}_i,~~ V^{-}_{phn,i}=(\gamma_{\mathrm{phn}}(1+n_B(\omega_0)))^{1/2} a_i$, where $n_B(\omega)$ is the Bose-Einstein distribution function.\cite{Ajisaka2015}

To numerically find the steady-state, the density matrix (of the full Fock many-exciton space) is stretched to a vector form, which allows us to write the master equation as 
$\dot{\vec{\rho}}=\hat{M} \vec{\rho}.$ The steady-state density matrix $\vec{\rho}_{SS}$ is thus defined as 
the null space of the superoperator $\hat{M}$. For the calculation of the ETC interacting with localized 
modes the density matrix was further projected onto a single-particle density matrix \cite{Pershin2008}. 

\subsection{Calculating heat currents}

The total energy of the system at steady state is simply defined as $E=\mathrm{Tr} \left( \rho_{SS} \cH \right)$. Since the Hamiltonian 
is time independent, it follows that $0=\frac{dE}{dt}=\mathrm{Tr} \left( -\hat{M}\rho_{SS} \cH \right)$. The structure of the 
Lindbladian, and specifically its separability into different environments, leads to the direct definition of the power input and output,   
$Q_{in}=\mathrm{Tr} \left( -\hat{M}_{source}\rho_{SS} \cH \right)$ and $Q_{out}=\mathrm{Tr} \left( -\hat{M}_{sink}\rho_{SS} \cH \right)$. 
For the {\sl local} heat currents,  since one has to define a local Hamiltonian, the situation is slightly more complex \cite{Wu2009}. We follow 
Ref.~\cite{Wu2009} and define the local Hamiltonian as $h_i=\epsilon_i c^{\dagger}_i c_i+\half t_{i,j} c^{\dagger}_i c_j+h.c$.  The total local heat current flowing 
through the site $i$ is then defined from $\frac{d}{dt}\langle h_i \rangle = -i \mathrm{Tr} \left( [h_i,\cH] \rho \right) -\mathrm{Tr} \left( h_i \hat{M} \rho \right)$. Since the total heat current
flowing to and from a site $i$ vanishes at steady state, this last expression can be written as $\frac{d}{dt}\langle h_i \rangle=\sum_j \dot{Q}_{i \leftrightarrow j}$, which allows us to identify the local heat currents.

\section{Power and efficiency}

\begin{center}
		\begin{figure}[h!]
			
		\includegraphics[width=16truecm,trim={0 2mm 0.6mm 0},clip]{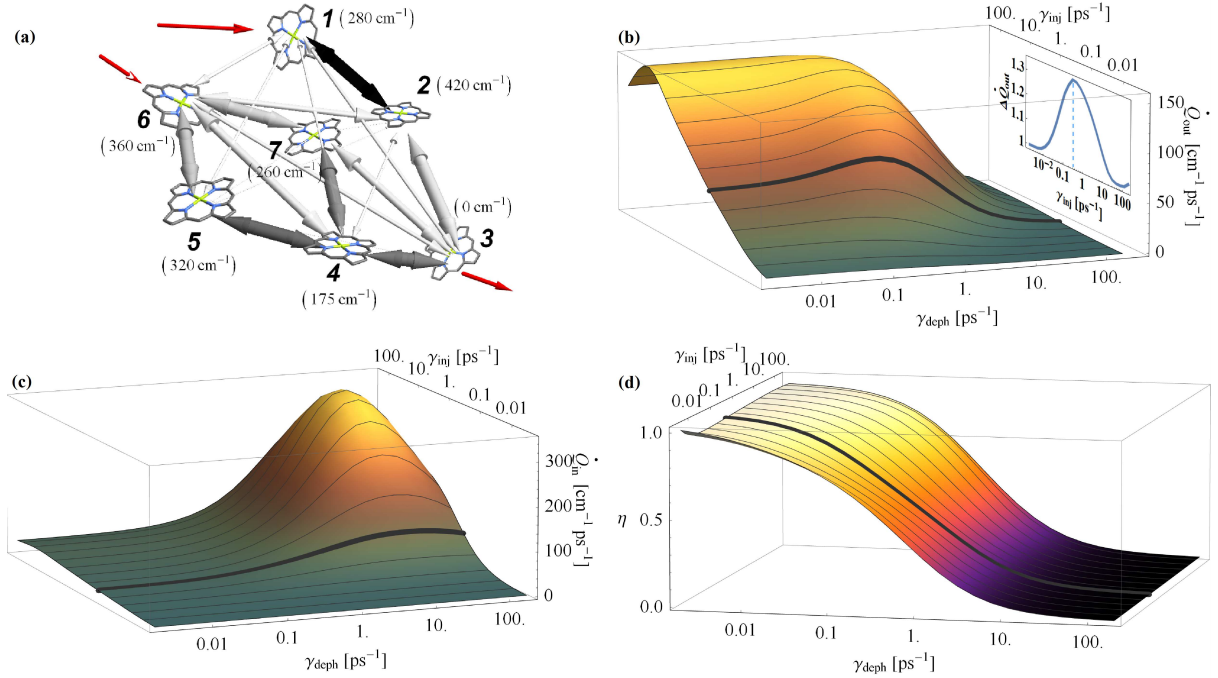}
			
			\caption{Energy flux and efficiency of the FMO exciton transfer complex. (a) Network structure of the FMO exciton transfer complex. Exciton energies are given in parentheses. Arrow thickness and shading correlate with interchromophore coupling strength. The network can be described roughly as two weakly connected sub-networks (branches). (b), Power output 
				$\dot{Q}_{\mathrm{out}}$ as a function of the energy injection rate $\gamma_{\mathrm{inj}}$ and dephasing rate $\gamma_{\mathrm{deph}}$. For any given $\gamma_{\mathrm{inj}}$ the power output exhibits 
				a maximum as a function of $\gamma_{\mathrm{deph}}$. The inset, which presents the ratio $\Delta \dot{Q}_{\mathrm{out}}$ between $\dot{Q}_{\mathrm{out}}$ at zero dephasing and $\dot{Q}_{\mathrm{out}}$ at 
				optimal dephasing, shows that for $\gamma_{\mathrm{inj}}\sim 0.5 \mathrm{ps}^{-1}$, an optimal increase in output of $\sim 30 \%$ was obtained (marked by a bold line in the main figure). (c), Power  input $\dot{Q}_{\mathrm{in}}$ (i.e., the power that the ETC is able to absorb from the antenna) as a function of $\gamma_{\mathrm{inj}}$ and $\gamma_{\mathrm{deph}}$. Surprisingly, $\dot{Q}_{\mathrm{in}}$ exhibits a maximum, demonstrating that the presence of dephasing increases the ability of the ETC to absorb energy from the antenna. (Bold line is the same as in (b)). {\bf d}, thermodynamic efficiency $\eta$ as a function of $\gamma_{\mathrm{inj}}$ and $\gamma_{\mathrm{deph}}$. The efficiency is monotonically decreasing with $\gamma_{\mathrm{deph}}$, regardless of $\gamma_{\mathrm{inj}}$.} 
			\label{fig1}
		\end{figure} 
	\end{center}

To begin, in Fig.~\ref{fig1}(b) the output power $\dot{Q}_{\mathrm{out}}$ is plotted as a function of the energy injection rate $\gamma_{\mathrm{inj}}$ and dephasing rate $\gamma_{\mathrm{deph}}$ (in the supplementary information we show that the power output and the exciton current are not proportional to each other, supporting our choice to study the power output). 
$\dot{Q}_{\mathrm{out}}$  exhibits non-monotonic behavior and displays a maximum at a given $\gamma_{\mathrm{deph}}$. Both the value of $\gamma_{\mathrm{deph}}$ at which the power is 
maximal and the optimal value of the power depend on $\gamma_{\mathrm{inj}}$. The inset of Fig.~\ref{fig1}(b) displays the ratio $\Delta \dot{Q}_{\mathrm{out}}$ between $\dot{Q}_{\mathrm{out}}$ at 
zero dephasing and $\dot{Q}_{\mathrm{out}}$ at optimal dephasing and shows an optimal increase in output of $\sim 30\%$ for $\gamma_{\mathrm{inj}}\sim 0.5 \mathrm{ps}^{-1}$. Interestingly, the optimal 
values of $\gamma_{\mathrm{deph}}$ and  $\gamma_{\mathrm{inj}}$ are of the same order, which is in accordance with previous observations on the conversion of time-scales for optimal performance \cite{Lloyd2011}. The optimal dephasing rate is such that quantum correlations are still present and important (as will be demonstrated in the following section), corresponding to the regime of environment-assisted quantum transport \cite{Rebentrost2009,Plenio2008}. Importantly, our results demonstrate that the power output benefits from the presence of dephasing even when the excitation is continuous incoherent. \cite{Pelzer2014,Manzano2013}

Fig.~\ref{fig1}c shows the power input $\dot{Q}_{\mathrm{in}}$, i.e. the power that the ETC is able to absorb from the antenna, as a function of $\gamma_{\mathrm{inj}}$ and 
$\gamma_{\mathrm{deph}}$. Surprisingly, $\dot{Q}_{\mathrm{in}}$ exhibits a maximum for some finite dephasing rate, implying that dephasing not only allows the transmission of more power by the ETC, it also facilitates its absorption from the antenna complex. To the best of our knowledge, this effect has not been discussed previously, since it cannot be observed by studying the dynamics of excited excitons in the absence of the antenna. The origin of this effect is the efficient depopulation of the sites coupled to the antenna. Evidence for this can be found looking at the the position of the maximal power input ,which is at higher dephasing rates compared to the maximum in the power output, and is correlated with the maximum in the exciton current (see inset of Fig.~\ref{fig2}a and supplementary material). 

We mention here that the nonmonotonic behaviors of the power input and output are robust, and persist when considering the full FMO trimer (including interactions between the individual monomers, with and without the eighth chromophore) and for the case of pulse-shaped excitations (in contrast to the continuous excitation considered here), both cases discussed in detail in the supplementary information. 

The ratio between $\dot{Q}_{\mathrm{out}}$ and $\dot{Q}_{\mathrm{in}}$ naturally defines the thermodynamic efficiency $\eta$, which is plotted as a function of $\gamma_{\mathrm{inj}}$ and 
$\gamma_{\mathrm{deph}}$ in Fig.~\ref{fig1}d. $\eta$ is monotonically decreases with $\gamma_{\mathrm{inj}}$, indicating that along with the increase in power absorption 
from the antenna and in the power output to the reaction center that accompany increasing $\gamma_{\mathrm{deph}}$, the heat flux dissipated into the local environment also increases. One can (rather intuitively) conclude from this outcome that if the finite dephasing time is a result of an evolutionary path chosen to optimize photosynthetic performance, the quantity to maximize was not the thermodynamic efficiency, but rather the power output to the reaction center. 

\section{Local heat currents}
What is the origin of this dephasing-induced enhancement of the power output? We address this question from the perspectives of heat currents and power output. To this end, we calculate the {\sl local} heat currents between the nodes (chromophores) of the ETC network for a constant injection rate of $\gamma_{\mathrm{inj}}=1 \mathrm{ps}^{-1}$ (see appendix for details). Fig.~\ref{fig2}a shows the local heat current 
$\dot{Q}_{i \leftrightarrow  j}$ as a function of the dephasing rate  $\gamma_{\mathrm{deph}}$ between selected pairs of sites $i$ and $j$. The sites corresponding to pairs  $\{ 1,6\}$ and $\{2,5\}$ reside on different branches of the exciton transfer network ("inter-branch"), and those corresponding to pairs $\{ 1,2\}$ and $\{6,7\}$ reside on the same branch ("intra-branch"). Consequently, dramatically different local heat current behaviors are observed: for intra-branch pairs, the heat current always flows in the same direction, while for inter-branch pairs it changes sign, i.e., the heat flow reverses direction.

Given that the "inter-branch" heat current changes sign as a function of
$\gamma_{\mathrm{deph}}$,  it must vanish at some value $\gamma_{\mathrm{deph},0}$ (marked by the arrow in Fig.~\ref{fig2}a). As is evident from the figure, both inter-branch currents vanish at the same dephasing rate  $\gamma_{\mathrm{deph},0}$. More importantly, $\gamma_{\mathrm{deph},0}$ ~is the same rate at which the power output is maximized. This is seen in the inset of Fig.~\ref{fig2}a, where the power output  $\dot{Q}_{\mathrm{out}}$ is plotted (solid line) as a function of   $\gamma_{\mathrm{deph}}$, the location of the maximum is marked by an arrow. Also plotted in the inset of Fig.~\ref{fig2}a is the exciton current, $J_{ex}$, which is in many cases used as a measure of efficiency. One can clearly see the difference in behavior between the heat current (which is a natural measure of performance) and the exciton current (more details are provided in the supplementary material).

Supported by the observation that the local inter-branch heat flow vanishes at maximum power output  suggests that the following scenario for dephasing-induced power enhancement takes place. As energy is injected into the ETC (from sites 1 and 6), it flows toward the reaction center (site 3). However, it also flows between the two branches and is prevented from reaching the reaction center. As the dephasing rate increases, the heat current, blocked from flowing between the branches, is {\sl directed} along the branches toward the reaction center, resulting in less dissipation and a higher power output. Put differently, enhancement of the dephasing rate forces the heat to flow to the reaction center along the shortest possible path, which results in a reduction in dissipation. This scenario differs from that described by the delocalization mechanism, which suggests that in the absence of dephasing, the excitons are strongly localized and only become delocalized when dephasing takes place.

  \begin{figure}[h]
  	
  	\includegraphics[width=8.5truecm]{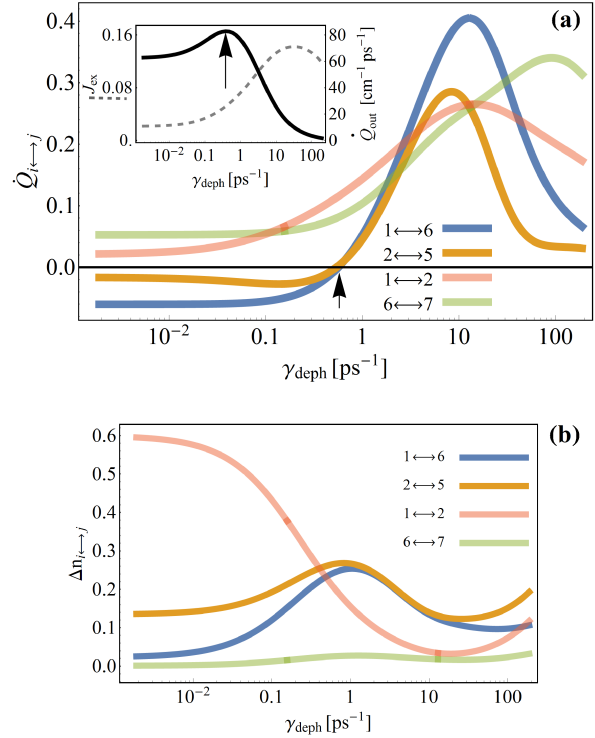}
  	
  	\caption{Local heat currents inside the FMO electron transfer complex. (a) local heat currents $\dot{Q}_{i \leftrightarrow j}$ between selected pairs of sites as a function of the 
  		dephasing rate $\gamma_{\mathrm{deph}}$. The arrow indicates the point $\gamma_{\mathrm{deph},0}$ at which the heat current between sites residing on different branches of the FMO complex (inter-branch currents) vanishes, while the intra-branch heat currents do not show any particular feature. The inset shows power output  $\dot{Q}_{\mathrm{out}}$ as a function of 
  		$\gamma_{\mathrm{deph}}$. The point at which the power output is maximal ( $\gamma_{\mathrm{deph},0}$, marked by an arrow) is also the point at which the inter-branch heat currents vanish. Also in the inset is the exciton current $J_{ex}$ (dashed gray line) emphasizing the difference between the heat and exciton currents. (b) Exciton occupation differences $\Delta n_{i \leftrightarrow j}$ as a function of $\gamma_{\mathrm{deph}}$ for the same selected pairs as in (a). Because the sign of the differences in exciton occupation does not change, it cannot be the origin of the change of sign in the local heat currents.  }
  	\label{fig2}
  \end{figure}

What could be the microscopic origin of this directed heat flow? One possible explanation is that energy diffuses between different sites, driven by differences in the local exciton 
occupation. If this were the case, a change of sign in the local energy current would imply a corresponding change of sign in exciton occupation differences. In Fig.~\ref{fig2}{\bf b} the differences 
$\Delta n_{i \leftrightarrow  j}$ in occupation for the same pairs of sites plotted in \ref{fig2}{\bf a} are plotted as a function of the dephasing rate. As can be observed, although the local heat currents vanish and change sign (for the intra-branch pairs), the differences in occupation do not change sign, indicating that the directed heat flow is not due to the simple diffusion of excitons. 

We therefore propose the following mechanism. Consider first a fully quantum system, i.e., with no dephasing at all (but in the presence of a source and sink). The non-equilibrium heat flux implies that at the source, the eigen-states of the system are excited, and between them they exchange energy that is eventually transferred to the sink. The real-space directions of the heat currents will be determined by the structures of the wave-functions and the eigen-energies and the weight of each wave-function on the local sites. Next, consider a system with the limit of a very high dephasing rate. In this case, quantum correlations vanish and the system behaves like a classical diffusive system  \cite{Dubi2009c}. 

The ETC then resembles a (thermal) resistor network, and the directions of the heat currents are 
determined by either the energy drop (i.e. heat will flow from high to low on-site energies) or by the direction of the overall heat current (note that the system is driven out of 
equilibrium by an external heat current). The directions of the local heat currents in these two limits can be different, and the transition between them tuned by the 
dephasing rate, leading to a vanishing local heat currents - and a resulting minimum dissipation and maximum power output - at some specific dephasing rate. 

\section{Insight from a simplified model}
\begin{figure}[h]
	\includegraphics[width=8.5truecm]{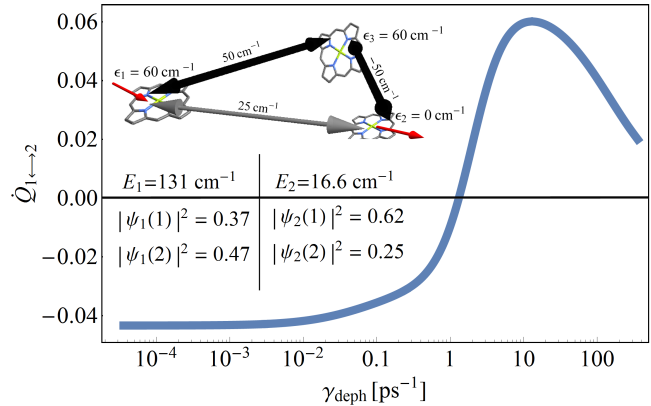}
	\caption{Local heat current of a simple exciton transfer network. The local heat current $\dot{Q}_{1 \leftrightarrow 2}$ between the sites 1 and 2 of the simplified ETC network (shown in the inset), as a function of the dephasing rate $\gamma_{\mathrm{deph}}$.   } 
	\label{fig3}
\end{figure}

Noting the rather complicated network structure of the FMO complex, it is useful to study the interplay between dephasing, structure and local heat currents, as described above, in a simpler system. To that end, we studied the simplest exciton transfer complex that displays directed heat flow. Schematically depicted in the upper inset of Fig.~\ref{fig3}, the complex comprised only three levels. Heat flows into and out of the network from sites 1 and 2, respectively, which are coupled to the third site non-symmetrically. The Hamiltonian for this complex can be written as  $\cH=\sum_{i=1}^3 c^{\dagger}_i c_i+\sum_{i,j} t_{i,j}c^{\dagger}_i c_j$, with $\epsilon_1=\epsilon_3\neq \epsilon_1$, and we choose $t_{1,2}=t_0,~t_{1,3}=t_1,~t_{2,3}=-t_1$. 

Due to the system's small size, the steady state currents can be found analytically, although the full expressions are too cumbersome to be presented here. Nonetheless, dephasing-induced directed heat flow can be observed by examining the limits of vanishingly small and large dephasing rates. For simplicity, we consider the case in which is the injection rate is small and the exciton injection and extraction rates are the same, and $t_1=t_0$. For the limit of vanishing dephasing, the local heat current between sites 1 and 2 is given by $Q_{1 \leftrightarrow 2}\approx -\frac{t_0 \gamma (\epsilon_1-\epsilon_2)(\epsilon_1-\epsilon_2+t_0)}{(\epsilon_1-\epsilon_2+t_0)^2+6t_0^2}$, which can be either negative or positive (i.e., flow can occur from site 1 to 2 or vice versa), depending on the value of $t_0$. On the other hand, for the limit of very large dephasing time, the local (diffusion-like) heat current is  $Q_{1 \leftrightarrow 2}\approx \frac{t_0 (\epsilon_1-\epsilon_2)}{\gamma_{\mathrm{deph}}}$, which is always directed along the drop in energy.

Further insight can be obtained via a numerical example of the same system for the parameters shown in the upper inset of Fig.~\ref{fig3} (the parameters were rather arbitrarily chosen to have similar values to the FMO complex, but the results that follow are rather general), which shows $Q_{1 \leftrightarrow 2}$ as a 
function of the dephasing rate $\gamma_{\mathrm{deph}}$. For weak dephasing, the heat flows against the drop in energy. As the dephasing rate increases, it reaches a certain value at which no heat flows between the sites. Finally, for large dephasing rates, the heat flow changes sign and flows along the energy drop. To better understand this observation, consider first the zero dephasing case, in which the ETC is adequately described by the Hamiltonian eigenstates and heat flows from the high to the low energy state. The numerical values for the eigen-energies and the weights of the corresponding wave functions are shown in the bottom inset of Fig.~\ref{fig3}. 
The localization of the high-energy state $\psi_1$ and the low energy state$\psi_2$ on sites 2 and 1, respectively, imply that a real-space heat current exists from site 2 to site 1. 
As the dephasing rate increases, the coherence between the real-space parts of the wave-functions are reduced and
the system becomes more and more diffusive, rendering the contribution made by diffusive heat flow 
increasingly important.  Eventually, it fully compensates the coherent heat flow, resulting in zero net heat flow between the sites. Further increase in the dephasing rate 
results in a local heat current dominated by the diffusive heat flow. 

\section{Interaction with a localized vibration} \label{secLocalized}
The model discussed above assumes a flat vibration spectrum characterized by a single parameter, the dephasing rate. However, because FMO chromophore vibrations possess a specific spectrum structure, it is of interest to test whether the above results also hold when taking such a structure into account. Specifically, we focus on the non-continuous part of the spectrum, which refers to the $180 \mathrm{cm}^{-1}$ vibrational mode of the chromophores \cite{Adolphs2006,Pachon2012}. To account in the model for this vibration, we add to the FMO Hamiltonian a term that includes the local vibration mode (at each chromophore) and an exciton-vibration interaction term, \beq \cH_{vib}=\hbar \omega_0 \sum_i a^\dagger_i a_i +\lambda \sum_i(a_i+a^\dagger_i) c^\dagger_i c_i ~~,\eeq where $a^\dagger (a)$ creates (annihilates) 
a vibration at chromophore $i$ with energy $\omega_0$, and $\lambda$ is an 
exciton-vibration interaction term, taken to be $\lambda=150$ cm$^{-1}$ \cite{Mourokh2014}. The interaction of the vibrations with the environment cause them to decay (described by an additional Lindblad term in the quantum master equation \cite{Ajisaka2015}, see appendix section), with a decay rate of $\gamma_{\mathrm{phn}}$ (inversely proportional to the vibration lifetime).  

\begin{center}
		\begin{figure}[h]
			
			\includegraphics[width=16.5truecm]{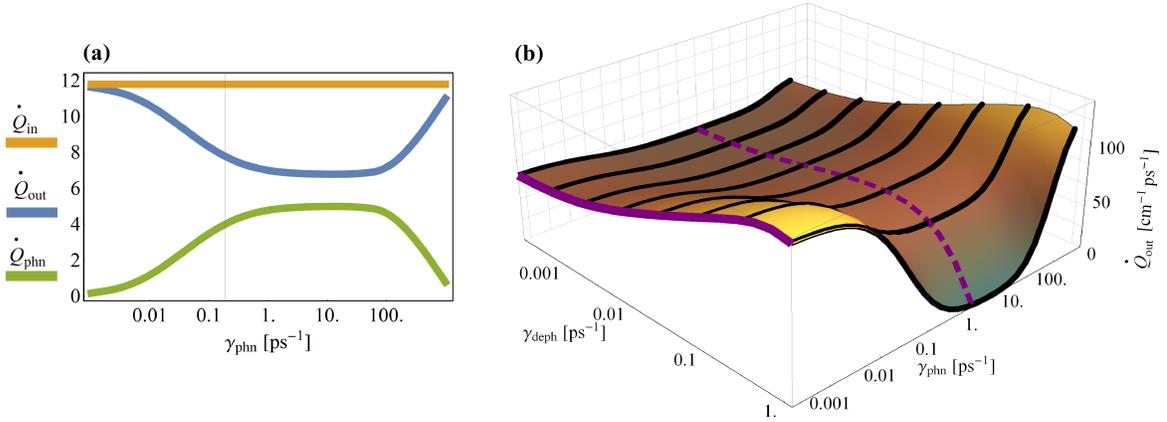}
			
			\caption{Interaction with a localized vibration. (a) Power input $\dot{Q}_{\mathrm{in}}$, power output $\dot{Q}_{\mathrm{out}}$, and power absorbed by the phonon mode 
				$\dot{Q}_{\mathrm{phn}}$, as a function of the phonon decay rate $\gamma_{\mathrm{phn}}$ and in the absence of dephasing. In contrast to the case of the continuous vibration spectrum (i.e., dephasing), the presence of localized vibration does not lead to a maximum in either the power input or output, only to their reductions. (b) power output $\dot{Q}_{\mathrm{out}}$ in the presence of both the localized vibration 
				and dephasing, as a function of $\gamma_{\mathrm{phn}}$ and $\gamma_{\mathrm{deph}}$. The presence of the localized mode hinders the power output, such that at resonance there is a two-fold reduction in the maximum power output (compare solid and dashed purple lines).}
			\label{fig4}
		\end{figure}
	\end{center}

In Fig.~\ref{fig4}a the power input and output ($\dot{Q}_{\mathrm{in}}$ and $\dot{Q}_{\mathrm{out}}$, respectively), and power absorbed by the localized vibration $\dot{Q}_{\mathrm{phn}}$ (dissipated to the vibrational environment) are plotted as a function of the vibration decay rate  $\gamma_{\mathrm{phn}}$ in the absence of dephasing. In contrast to the case of a continuous vibrational spectrum, the power input is constant and does not show any change. This outcome is due to the localization of the vibration on the chromophore and the absence of energy transfer (due to exciton-vibration coupling) between the ETC eigenstates (see supplementary information for a more detailed discussion). The power output $\dot{Q}_{\mathrm{out}}$ decreases (vs. increases in the case of dephasing) until it reaches a minimum value when the decay rate is in resonance with the heat injection rate. As the decay rate increases further, the power output increases, returning to its original value, because the vibration is so short-lived that it can no longer absorb energy from the exciton.

In Fig.~\ref{fig4}b, the power output $\dot{Q}_{\mathrm{out}}$ in the presence of both localized vibration and dephasing is plotted as a function of  $\gamma_{\mathrm{phn}}$ and  $\gamma_{\mathrm{deph}}$. Although the non-monotonic behavior of  $\dot{Q}_{\mathrm{out}}$ with  $\gamma_{\mathrm{deph}}$ is maintained, it is influenced by the localized vibration, which shifts the position of the maximum and reduces the maximum value of the power output. Comparing the case of the vanishing vibration decay rate (which is similar to the absence of vibration, solid purple line) with that of vibrational resonance (dashed purple line), we observe an approximately two-fold reduction in the maximal power output, and the remaining heat is dissipated into the localized vibration.

\section{Dephasing and durability}
Besides its benefit to exciton transport, quantum coherence has also been suggested to contribute to transport stability, i.e., the robustness of transport properties to changes in the local parameters  \cite{Shabani2014}. From this perspective, therefore, we test whether the presence of dephasing preserves power output durability in the face of drastic changes in FMO structure. To this end, we define the quantity $R_{i\leftrightarrow j}$ as follows: for selected sites $i$ and $j$ in the FMO complex, $R_{i\leftrightarrow j}$ is the ratio between the power output of the full FMO complex and that when the link connecting $i$ and $j$ is removed (i.e., to mimic a damaged ETC). The closer $R_{i\leftrightarrow j}$ is to unity, the more stable the FMO complex and the greater its resilience to damage.

\begin{figure}[h!]
 	\includegraphics[width=8.5truecm]{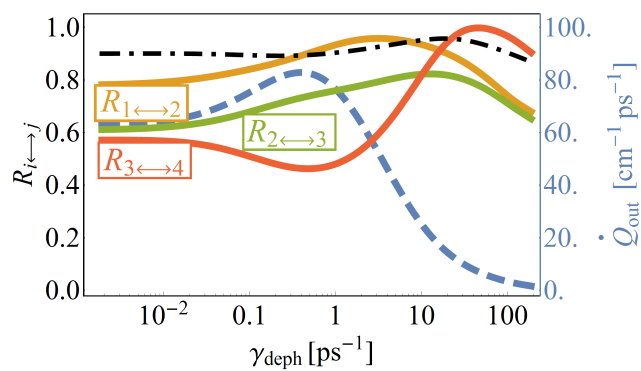}
 	\caption{Stability of power output against exciton network damage. Ratio $R_{i\leftrightarrow j}$ between the power output of the full FMO complex and that when the link connecting $i$ and $j$ is removed, for selected pairs, as a function of the dephasing rate. Dephasing increases the resilience to damage in the  $1\leftrightarrow 2$ branch of the FMO complex but seems to decrease the durability of the other branch. The dotted-dashed line, which represents the ratio $R_{i\leftrightarrow j}$ averaged over all pairs, indicates that the overall effect of dephasing on resilience is small. The power output is plotted (blue dashed line) for comparison. }
 	\label{fig5}
 \end{figure}

In Fig.~\ref{fig5},  $R_{i\leftrightarrow j}$ is plotted for the pairs $1\leftrightarrow 2,~2\leftrightarrow 3, 3\leftrightarrow 4$ as a function of  $\gamma_{\mathrm{phn}}$. For comparison, the dashed blue line is the power output of the undamaged complex (same as the inset of Fig.~\ref{fig2}). The result, however, is inconclusive: for damage to the right branch of the FMO network (sites 1 and 2), dephasing seems to be beneficial (i.e., to enhance R or, equivalently, to reduce the effect of the removal of the  $i\leftrightarrow j$ connection on the power output). In the left branch of the FMO network, in contrast, dephasing reduces $R_{3\leftrightarrow 4}$, making the system more susceptible to damage. Qualitatively similar results are obtained if the coupling through the $i \leftrightarrow j$ link is only reduced (rather then fully removed). The dot-dashed black line in Fig.~\ref{fig5} shows the $R_{i\leftrightarrow j}$  averaged over all links of the FMO network. Thus, it seems that the effect of dephasing on the resilience of the network to damage is marginal. Reflecting this result on the evolution of the FMO complex, we arrive at the following hypothesis: the two-branch geometry of the FMO complex has evolved to enhance resilience and stability \cite{Hoyer2010,Cheng2006} (as clearly a single-branch network is far more susceptible to damage). The internal structure of the FMO complex (which in turn determines the dephasing time and the amount of quantum coherence) has evolved to provide maximal power output within confines of a two-branch geometry. Testing this hypothesis (for instance via genetic numerical algorithms mimicking the evolution of the FMO complex) are currently underway.

\section{Concluding remarks}

In this work, we studied energy transfer in the FMO photosynthetic exciton transfer complex from the perspective of steady-state heat transport. This approach illuminated the connection between dephasing, power output, local heat currents and network geometry, and it showed the benefit of exploiting coexisting quantum and classical correlations to facilitate exciton transfer in photosynthetic systems. Similar calculations including more complicated geometries, environments with a specific spectral density, and environments out of equilibrium and beyond the Markov approximation are currently underway. We hope that the results presented here will be the impetus behind new experiments that directly probe the power output and energy transfer in photosynthetic excitonic transfer complexes.


\begin{acknowledgement}
This work was partially supported by the Adelis foundation.
\end{acknowledgement}

\providecommand*\mcitethebibliography{\thebibliography}
\csname @ifundefined\endcsname{endmcitethebibliography}
{\let\endmcitethebibliography\endthebibliography}{}


\end{document}